\documentclass{article}
\pdfoutput=1
\usepackage{spconf,amsmath,graphicx}
\usepackage{mathtools}
\usepackage{amssymb}
\usepackage{amsthm}
\usepackage{thmtools}
\usepackage{times}
\usepackage{epsfig}
\usepackage{graphicx}
\usepackage{amsmath}
\usepackage{amssymb}
\usepackage{booktabs}
\usepackage{hyperref}
\usepackage{color}
\usepackage{multirow}
\usepackage{graphicx}
\usepackage{multirow}
\usepackage{graphicx,multirow}
\usepackage{wrapfig}
\usepackage{subcaption}
\usepackage{tikz} 
\usepackage{pgfplots}
\usetikzlibrary{fit}
\pgfplotsset{compat=newest}%
\usepackage{enumitem}
\usepackage[accsupp]{axessibility}  
\usepackage{enumitem}
\setlist{nosep, leftmargin=14pt}

\usepackage{mwe} 

\title{Medical Image Segmentation Using Directional Window Attention}
\name{Daniya~Najiha~A.~Kareem$^1$, Mustansar~Fiaz$^1$, Noa~Novershtern$^2$, and Hisham~Cholakkal$^1$ 
}
\address{$^1$Mohamed bin Zayed University of Artificial Intelligence, UAE \hspace{1.5mm} $^2$Weizmann Institute of Science, Israel}
\begin{document}

%
\maketitle
\begin{abstract}
Accurate segmentation of medical images is crucial for diagnostic purposes, including cell segmentation, tumor identification, and organ localization. Traditional convolutional neural network (CNN)-based approaches 
struggled to achieve precise segmentation results due to their limited receptive fields, particularly in cases involving multi-organ segmentation with varying shapes and sizes. The transformer-based approaches address this limitation by leveraging the global receptive field, but they often face challenges in capturing local information required for pixel-precise segmentation.   In this work, we introduce DwinFormer, a hierarchical encoder-decoder architecture for medical image segmentation comprising a directional window (Dwin) attention and global self-attention (GSA) for feature encoding. The focus of our design is the introduction of Dwin block within DwinFormer that effectively captures local and global information along the horizontal, vertical, and depthwise directions of the input feature map by separately performing attention in each of these directional volumes.  To this end, our Dwin block introduces a nested Dwin attention (NDA) that progressively increases the receptive field in horizontal, vertical, and depthwise directions and a convolutional Dwin attention (CDA) that captures local contextual information for the attention computation.  
While the proposed Dwin block captures local and global dependencies at the first two high-resolution stages of  DwinFormer, the GSA block encodes global dependencies at the last two lower-resolution stages. Experiments over the challenging 3D  Synapse Multi-organ dataset and Cell HMS dataset demonstrate the benefits of our DwinFormer over the state-of-the-art approaches. Our source code will be publicly available at \url{https://github.com/Daniyanaj/DWINFORMER}. 
\end{abstract}


\section{Introduction}
\label{sec:intro}
Medical image segmentation is a challenging task that requires pixel (voxel)-precise localization of cells, tumors, and human organs for diagnostic purposes \cite{nnunet} \cite{a} \cite{b}. This challenging task is generally addressed using a U-Net \cite{ronneberger2015u}, where the encoder creates a low-dimensional representation of the input 3D image, and the decoder maps it to an accurate segmentation mask.
Previous CNN-based methods \cite{ronneberger2015u} \cite{3dunet} \cite{nnunet} struggled to achieve accurate segmentation results due to their limited receptive field. Various efforts have been made using dilated convolution \cite{zhang2020inter}, feature pyramid \cite{gridach2021pydinet}, contextual attentions \cite{wang2019volumetric, fang2020spatial} to handle the long-range dependencies.
Nevertheless, these approaches still limit their learning capabilities due to 
the locality of the receptive fields. This may lead to sub-optimal
segmentation, especially where structures of the tissues are variable in shape and scale.
Recently, transformers have achieved state-of-the-art performance on 3D medical image segmentation tasks by employing a self-attention (SA) mechanism for capturing long-range dependencies. 
Transformers are remarkably good at capturing global interactions within images, leading to a larger receptive field and more precise predictions. 
Consequently, there has been a surge in the development of transformer-based methods and hybrid models combining CNNs and transformers, which have substantially enhanced segmentation accuracies \cite{transunet, nnformer}. 

The Swin transformer \cite{liu2021swin} has recently emerged as a promising solution for  3D segmentation \cite{nnformer} tasks due to the effective extraction of global and local dependencies. It utilizes non-overlapping window-based multi-head attention, which has a linear complexity advantage over the quadratic complexity found in ViTs 
\cite{dosovitskiy2020image}. However, the Swin transformer still has limitations in explicitly encoding global interactions due to its limited attention area which is restricted inside the window. Increasing window size or applying attention to full resolution would lead to higher computational costs and a parameter-heavy model. To this end, we propose our DwinFormer which strives to 
effectively capture local and global information while limiting model complexity.

\noindent\textbf{Contribution:} 
In this work, we propose a hierarchical encoder-decoder architecture, dubbed as DwinFormer for medical image segmentation tasks. Our DwinFormer comprises directional window (Dwin) attention and global self-attention (GSA) blocks. 
The objective of our novel Dwin attention block is to encode the local and global representations across horizontal, vertical, and depthwise directions. Specifically, we introduce nested Dwin attention (NDA) and convolutional Dwin attention (CDA) within the Dwin block to progressively enlarge the receptive field across horizontal, vertical, and depthwise directions as well as to encode the local contextual information, respectively. 
%
Experimental results over 3D multi-organ Synapse (human organs) and  Cell HMS (microscopic zebrafish cells) segmentation datasets show the superiority of our method over state-of-the-art methods. 
\begin{figure*}[h]
\centering
\includegraphics[width=0.8\linewidth]{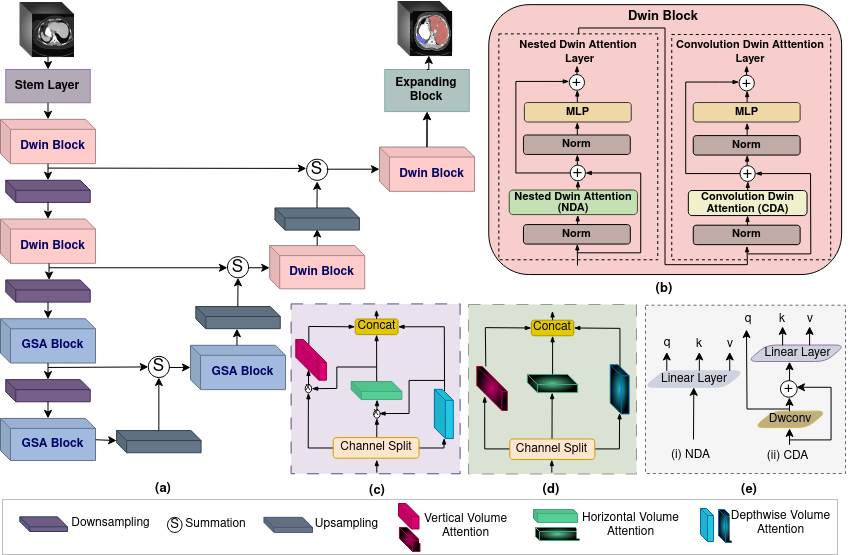}
   \caption{\textbf{(a)} Overall architecture of the proposed  DwinFormer having a hierarchical encoder-decoder framework.  
  In the encoder, the stem features are input to the  directional window (Dwin) block  to explicitly learn the local and global dependencies at high resolution in the initial two stages of the encoder, whereas global self-attention (GSA) block is applied in the later two stages to capture the global information.  In the decoder, the features are first upsampled and then added with the encoder features using a skip connection. 
  The focus of our design is the introduction  of \textbf{(b)} Dwin block into DwinFormer, enabling the effective capturing of local and global information in multiple directions within the input feature map. 
  The Dwin block consists of two components:\textbf{(c)}  nested Dwin attention (NDA) that gradually expands the receptive field in the depthwise, horizontal and vertical  directions,  and \textbf{(d)} convolutional Dwin attention (CDA) that strives to capture local contextual information  using depthwise convolution during the attention computation,. \textbf{(e)} shows the qkv computation for attention in (i) Nested Dwin Attention (NDA) (ii) Convolutional Dwin Attention (CDA). The NDA employs a  linear layer to obtain qkv while CDA additionally captures local information using a depthwise convolution.}
 \label{fig:overall_proposed_framework}
 \vspace{-0.4cm}
\end{figure*}

\section{Methods}

\label{sec:Meth}
\noindent\textbf{Motivation:} 
As previously mentioned, transformer-based and hybrid approaches employ self-attention operations, which require high computational costs. An example of this limitation is nnFormer model \cite{nnformer}, which incorporates  Swin transformer blocks. The Swin transformer has a restricted attention area, making it challenging to explicitly encode global interactions. Increasing window size or employing attention on full resolution would result in increased computational costs and a model with a large number of parameters. In addition, accurately predicting target boundaries remains a challenge. Therefore, we propose the importance of learning boundary regions within both local and larger spatial contexts. Our approach focuses on explicitly capturing local and global dependencies using high-resolution features, while also incorporating global dependencies from lower-resolution features. By doing so, we aim to enhance the associations among volumetric feature representations, leading to improved predictions of boundary regions.
\vspace{-0.2cm}
\subsection{Overall Architecture} 
\vspace{-0.1cm}
The overall architecture of the proposed method, dubbed as DwinFormer, is shown in figure \ref{fig:overall_proposed_framework}-a. Our model follows an encoder-decoder framework with varying resolutions at each stage for accurately capturing local and global dependencies. The 3D image is input to the stem layer to generate stem features which are downsampled and input to different stages of the encoder. Similarly, the low-resolution features are upsampled and input to the decoder stages, and finally the expanding block to output the final mask. 
The primary objective is to learn the diverse shapes of target regions by explicitly capturing both local and global dependencies. Specifically, the framework captures local and global feature dependencies using the directional window (Dwin) block  at the first two of the encoder stages and the last two decoder stages, all of which have high feature resolutions. The remaining encoder and decoder stages, which have relatively lower feature resolutions, are responsible for capturing global feature dependencies using a global self-attention (GSA) block. 

\subsection{Directional Window (Dwin) Block:}
As mentioned previously, incorporating a self-attention mechanism is crucial in accurately segmenting regions with diverse shapes and sizes, as it enables the network to capture long-range dependencies. However, solely relying on self-attention may limit the ability to learn local contextual information, as global features become dominant and can result in increased computational costs. Therefore, we present a novel Dwin attention block that performs separate attention across all dimensional volumes to better learn the local and global representations with enhanced underlying attention areas.

Our proposed Dwin block comprises a nested Dwin attention (NDA)  and a convolution Dwin attention (CDA) layers as shown in figure \ref{fig:overall_proposed_framework}-b.  In both Dwin attention layers, the number of input channels is divided to perform attention along horizontal, vertical, and depthwise volumes. 
Suppose $\mathcal{F} \in  \mathcal{R}^{H\times W\times D\times C}$ 
be the input, where $N=(H, W, D)$ represents the size of the 3D input (volume) and $C$  denotes the number of channels.  In each layer of the Dwin block, the horizontal volume attention is performed on windowed volumes ($H \times sw \times sd$), vertical volume attention is performed on windowed volume ($sh \times  W  \times sd$), and depthwise volume attention is performed on the volume of size  ($sh \times  sw  \times D$). Here $s$ refers to the number of windowed volumes  along a particular direction whereas $h$, $w$, and $d$ refer to the window dimensions which can be adjusted accordingly.   The Dwin attention block operations can be summarized as:
\begin{align}
\begin{split}
      \scriptsize  {\mathcal{\hat{F}}=NDA(\text{Norm} (\mathcal{F} )+\mathcal{F}},\quad
   \scriptsize {\mathcal{\bar{F}}= MLP(\text{Norm} (\mathcal{\hat{F}})+\mathcal{\hat{F}}},\\
  \scriptsize  {\mathcal{\hat{\hat{F}}}=CDA(\text{Norm} (\mathcal{\bar{F}} )+\mathcal{\bar{F}}},\quad
  \scriptsize  {\mathcal{\bar{\bar{F}}}= MLP(\text{Norm} (\mathcal{\hat{\hat{F}}})+\mathcal{\hat{\hat{F}}}},
    \end{split}   
 \end{align} 
\vspace{-0.1cm} 
where $\hat{F}$ and  $\bar{F}$ represent the intermediate and final features of the NDA layer, whereas  $\hat{\hat{F}}$ and  $\bar{\bar{F}}$ denote the intermediate and final features of the CDA layer.\\


\vspace{-0.3cm}

\noindent\textbf{Nested Dwin Attention (NDA) Layer}: 
The objective of nested Dwin attention layer is to increase the receptive field while reducing exponential computational complexity of standard self-attention for volumetric input. In addition, the proposed NDA layer captures the global dependencies for better segmentation. The fundamental attention operation of the NDA layer is shown in figure \ref{fig:overall_proposed_framework}-c, 
which focuses on increasing the attention area and improving the representations by addressing the dependencies across these divided channel sets. In this attention mechanism, we consecutively multiply the attention maps from right to left and finally, concatenate these attention maps as shown in figure \ref{fig:overall_proposed_framework}-c.  
\\
\noindent\textbf{Convolution Dwin Attention (CDA) Layer}: We also propose a CDA layer intending to encode the local contextual information as shown in figure \ref{fig:overall_proposed_framework}-d. To do so, we use a 5$\times$5 depthwise convolution during the computation of qkv features, as shown in figure \ref{fig:overall_proposed_framework}-e-ii, with an objective to encode the local contextual information to handle the issue of smaller patch sizes with an increased spatial context.

\subsection{Global Self-Attention Block}
\vspace{-0.1cm}
Our global self-attention block consists of two sequential standard self-attention (SA) layers,  which is responsible for capturing global dependencies at lower-resolution features. The 3D input (volume)  tensor $\mathcal{F} \in  \mathcal{R}^{H\times W\times D\times C}$   is first reshaped to a vector  $f \in  \mathcal{R}^{N\times C}$ and input to GSA block for global representation. 
Later, the output of the GSA block is reshaped back to the original 3D tensor.

\begin{table}[t]
    \begin{center}
    \scalebox{0.7}{
        \begin{tabular}{|l|c|c|}
        \hline
       Method    &DSC  & HD95 \\ \hline \hline
        U-Net \cite{ronneberger2015u} & 76.85 & - \\ 
       
        ViT \cite{dosovitskiy2020image}+CUP \cite{transunet}                  & 67.86 & 36.11 \\ 
        TransUNet \cite{transunet}      & 77.48 & 31.69 \\ 
        Swin-UNet \cite{swin-unet}       & 79.13 & 21.55 \\ 
        LeVit-UNet-384s \cite{levit}  & 78.53 & 16.84 \\ 
        MissFormer \cite{miss}   & 81.96 & 18.20 \\ 
        UNETR \cite{unetr}                    & 79.56 & 22.97 \\ 
        Axial Deeplab \cite{deeplab}           & 85.37   &9.16    \\ 
        nnFormer \cite{nnformer}           & 86.57   &10.63     \\ 
        \textbf{DwinFormer (Ours)}        & \textbf{87.38}    & \textbf{8.68}     \\ \hline
        \end{tabular}}
	\caption{Comparison with other state-of-the-art methods over multi-organ Synapse dataset. We report the mean of our results for 3 runs.  The best results are in bold.}
 \vspace{-0.5 cm}
	\label{all}
	\end{center}
\end{table}

\begin{table}
\centering
\scalebox{0.5}{
\begin{tabular}{|l|c|c|c|c|c|c|c|c|}
\hline
          & ARE            & VOI\_split    & Avg JI        & Avg DSC       & JI\textgreater{}70\% & DSC\textgreater{}70\% & JI\textgreater{}50\% & DSC\textgreater{}50\% \\ \hline
UNet \cite{3dunet}   & 0.44           & 1.42                 & 45.7          & 58.10          & 26.2                 & 41.2                  & 45.8                 & 65.7                  \\ 
nnFormer \cite{nnformer} & 0.41           & 1.17           & 52.5          & 64.01         & 39.3                & 53.5                 & 56.09                & 73.6                  \\ 
Swin-UNet\cite{swin-unet} & 0.41 & 1.17         & 53.4          & 65.09         & 40.0                & 55.9                  &  \textbf{59.8}                & 76.0                    \\ 
Ours      & \textbf{0.38}           & \textbf{1.09}         & \textbf{54.2} & \textbf{65.8} & \textbf{40.6}          & \textbf{56.11}         & 58.4        & \textbf{78.18}         \\ \hline
\end{tabular}}
\caption{State-of-the-art comparison on HMS dataset \cite{cellseg} dataset. We report the results in terms of ARE, $VOI_{split}$, overall accuracy (JI and DSC), and cell count accuracy (JI/DSC greater than 50\% or 70\%) metrics. 
We report the mean for 3 runs. 
The best results are in bold.}
\label{res}
\end{table}

\begin{table*}[t]
\centering
\scalebox{0.65}{
\begin{tabular}{|c|cc|cc|cc|cc|cc|cc|cc|cc|cc|}
\hline
\multicolumn{1}{|l|}{\multirow{2}{*}{\textbf{Method}}} & \multicolumn{2}{c|}{\textbf{Average}}       & \multicolumn{2}{c|}{\textbf{Aorta}}         & \multicolumn{2}{c|}{\textbf{Gall Bladder}}  & \multicolumn{2}{c|}{\textbf{Kidney(L)}}     & \multicolumn{2}{c|}{\textbf{Kidney(R)}}     & \multicolumn{2}{c|}{\textbf{Liver}}        & \multicolumn{2}{c|}{\textbf{Pancreas}}     & \multicolumn{2}{c|}{\textbf{Spleen}}        & \multicolumn{2}{c|}{\textbf{Stomach}}      \\ \cline{2-19} 
                         & \multicolumn{1}{c|}{DSC}    & HD95    & \multicolumn{1}{c|}{DSC}    & HD95    & \multicolumn{1}{c|}{DSC}    & HD95    & \multicolumn{1}{c|}{DSC}    & HD95  & \multicolumn{1}{c|}{DSC}    & HD95    & \multicolumn{1}{c|}{DSC}    & HD95   & \multicolumn{1}{c|}{DSC}    & HD95   & \multicolumn{1}{c|}{DSC}    & HD95    & \multicolumn{1}{c|}{DSC}    & HD95   \\ \hline
UNETR \cite{unetr}   & \multicolumn{1}{c|}{79.50} & 22.97 & \multicolumn{1}{c|}{89.9} & 5.48 & \multicolumn{1}{c|}{60.55} & 28.69 & \multicolumn{1}{c|}{85.66} & 17.76 & \multicolumn{1}{c|}{84.80} & 22.44 & \multicolumn{1}{c|}{94.45} & 30.40    & \multicolumn{1}{c|}{59.24} & 15.82 & \multicolumn{1}{c|}{87.8} & 47.12 & \multicolumn{1}{c|}{73.99} & 16.05 \\ 

                         
nnFormer \cite{nnformer}    & \multicolumn{1}{c|}{86.57} & 10.63 & \multicolumn{1}{c|}{92.04} & 11.38 & \multicolumn{1}{c|}{70.17} & 11.55 & \multicolumn{1}{c|}{86.57} & 18.09 & \multicolumn{1}{c|}{86.25} & 12.76 & \multicolumn{1}{c|}{\textbf{96.84}} & \textbf{2.00}    & \multicolumn{1}{c|}{\textbf{83.35}} & \textbf{3.72} & \multicolumn{1}{c|}{90.51} &\textbf{16.92} & \multicolumn{1}{c|}{\textbf{86.83}} & 8.58 \\ 
\textbf{Ours} & \multicolumn{1}{c|}{\textbf{87.38}} & \textbf{8.68}  & \multicolumn{1}{c|}{\textbf{92.63}} & \textbf{4.56} & \multicolumn{1}{c|}{\textbf{73.17}} & \textbf{6.92}  & \multicolumn{1}{c|}{\textbf{87.64}} & \textbf{13.93}  & \multicolumn{1}{c|}{\textbf{87.41}} & \textbf{9.10}  & \multicolumn{1}{c|}{96.45} & 2.94 & \multicolumn{1}{c|}{83.11} & 4.23 & \multicolumn{1}{c|}{\textbf{92.17}} & 19.5 & \multicolumn{1}{c|}{86.49} & \textbf{8.24} \\ \hline
\end{tabular}}
\caption{Organ-wise segmentation comparison over Synapse Multi-organ dataset. The best results are in bold.}
\vspace{-0.4 cm}
\label{tb_nnunet}
\end{table*}
\vspace{-0.2cm}
\section {Experiments}
\vspace{-0.2cm}
\subsection{Datasets and Training Setup}
\vspace{-0.2cm}
\noindent\textbf{Multi-organ Synapse dataset:} is a multi-organ dataset,
includes 30 abdominal CT scans with 18 train and 12 validation scans.  It contains the segmentation task for eight organs including the liver, right kidney, left kidney, pancreas, gall bladder, stomach, spleen, and aorta. 

\noindent\textbf{Cell HMS dataset:} The HMS dataset was constructed, at Harvard Medical School \cite{cellseg}, from zebrafish cells containing 3 target categories. 
It contains 36 images with a resolution of 181×331×160. Out of which, we used 88\%   for training and 12\%  for testing. 

\noindent\textbf{Training Setup:}
The method is implemented in  PyTorch 1.8.0 and trained on an NVIDIA A100 GPU.
For the multi-organ synapse dataset, we adopt the pre-processing,  augmentation strategies, and training recipe from nnFormer \cite{nnformer}. 
We set the batch size to 2 and the initial learning rate to 0.01 and utilized a poly decay strategy to adjust the learning rate.
We set the momentum and weight decay as 0.99 and 3e-5 with SGD as the default optimizer. The training was done on 1000 epochs. We use a combination of soft dice loss and cross-entropy loss for training the network. 
For the Cell HMS dataset,   we set a learning rate of 0.0001, batch-size 5, and crop size  $128 \times 128 \times 128$. Following \cite{cellseg}, we adopt Adam optimizer and Dice loss with weights for training. 

\begin{figure}[t]
\begin{center}
\vspace{-0.2cm}
\includegraphics[width=\linewidth]{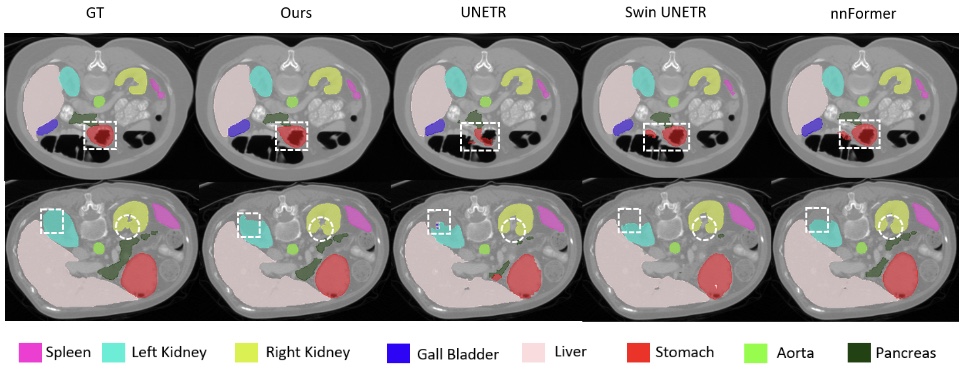}

\end{center}
\vspace{-0.5cm}
   \caption{Qualitative analysis on multi-organ synapse dataset \cite{landman2015miccai} shows that our method provides improved segmentation by accurately
detecting the organs with clear boundaries.} 
 \vspace{-0.5cm}
\label{fig:vis}
\end{figure}

\begin{figure}
\begin{center}
\includegraphics[width=0.88\linewidth]{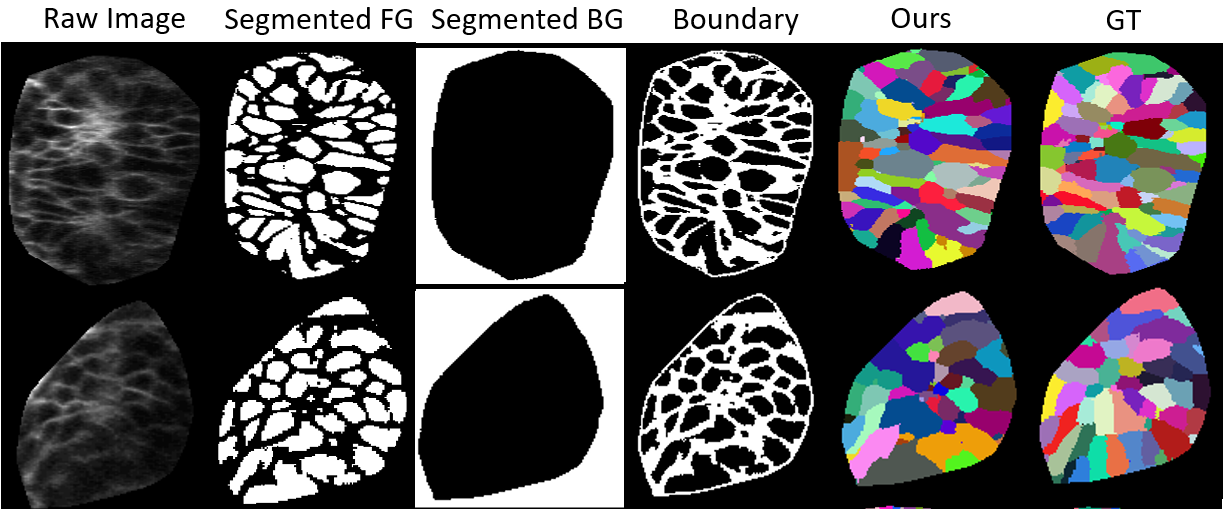}
\end{center}
\vspace{-0.5cm}
   \caption{Qualitative results of DwinFormer on the cell HMS dataset \cite{cellseg}. Rows correspond to views from different planes. DwinFormer predicts  foreground (FG),  background (BG), and cell boundary regions (columns 2-4)  which are post-processed to segment cell  instances (column~5)}
\label{fig:vishms}
\end{figure}

\begin{table}[t]
\centering
\scalebox{0.5}{
\begin{tabular}{|c|c|c|c|c|c|}
\hline
   & \textbf{Attention 1} & \textbf{Attention 2} & \textbf{Attention 3} & \textbf{DSC} & \textbf{HD95} \\ \hline
1. & Vertical            & Horizontal           & Depthwise            &     87.33       &        8.71       \\ \hline
2. & Depthwise             &    Vertical        &    Horizontal         &     87.09       &        8.82   \\ \hline
3. & Horizontal            &    Depthwise      &    Vertical         &     87.14        &         9.19     \\ \hline
4. & Horizontal            &  Vertical           &    Depthwise         &     87.21         &        9.01     \\ \hline
5. & Depthwise            &  Horizontal           &    Vertical        &     \textbf{87.38}   &  \textbf{8.68}  \\ \hline
6. & Vertical            &  Depthwise           &    Horizontal         &     87.06    &        9.12      \\ \hline
 
\end{tabular}}
\caption{Ablation studies on the selection of the order of depthwise, vertical, and horizontal volumetric attention mechanisms within nested Dwin attention (NDA) layer. 
The best results are in bold.}
\vspace{-0.2cm}
\label{tbl:order_within_Dwin_block}
\end{table}

\subsection{Comparison with state-of-the-art methods}
\label{sec:Res}


\noindent\textbf{Multi-organ Synapse Dataset:} Table \ref{all} shows our approach obtains a significantly higher dice score of  87.38\% and a better HD95 score of 8.68 multi-organ synapse dataset compared to existing methods. 
Furthermore, in Table \ref{tb_nnunet}, we conduct a detailed performance analysis and show that our method has significant improvement in terms of dice and HD95 scores in most of the organs. {The segmentation accuracies of organs like the left kidney, right kidney, and gall bladder shows considerable improvement which proves the efficiency of our method in segmenting complex boundaries. 
 The qualitative comparison in figure \ref{fig:vis} shows that our approach performs better compared to other methods in avoiding false segmentations and preserving the organ boundaries. \\
\noindent\textbf{Cell HMS Dataset:} Table \ref{res} shows that DwinFormer obtains improved performance compared to CNN-based and transformer-based methods for microscopic cell dataset. The qualitative comparison of our DwinFormer with groundtruth is demonstrated in figure \ref{fig:vishms} which shows that segmented output has a close resemblance to the actual segments.

\begin{table}[t]
\centering
    
    \scalebox{0.5}{
	\begin{tabular}{|c|l|l|l|l|l|l|}
        \hline
       Sr. No. & \multicolumn{1}{c|}{Stage 1} &  \multicolumn{1}{c|}{Stage 2} &  \multicolumn{1}{c|}{Stage 3} & Stage 4 &DSC & HD95    \\ \hline \hline
    1  & GSA block   & GSA block       & GSA block & GSA block& 81.90  &  16.02 \\ \hline
        2   &  Swin block   & Swin block     & Swin block       & Swin block & 86.57   &  10.63   \\ \hline
        3     & CSwin block  & CSwin block & CSWin block  &  CSWin block &  86.80 &  9.63 \\ \hline
        4     & Dwin block  & Dwin block        & Dwin block       & Dwin block & 86.98    & 9.11  \\ \hline
        5   & Dwin block & Dwin block   & GSA block       & GSA block      & \textbf{87.38} &  \textbf{8.68}   \\ \hline
             
        \end{tabular}}
	\caption{{Comparison of various variants of context aggregator blocks incorporating standard self-attention \cite{dosovitskiy2020image} as GSA,  Swin Transformer \cite{liu2021swin}, CSwin Transformer \cite{cswin} and the proposed Dwin block at different stages over multi-organ Synapse dataset.}}
	\label{ablation}
  
\end{table}


\vspace{-0.3cm}
\subsection{Ablation Study}
\vspace{-0.2cm}
We perform an ablation study over the multi-organ Synapse dataset to validate the effectiveness of our method. Firstly, we validate the order of horizontal, vertical, and depthwise attentions inside the nested Dwin attention (NDA) layer as in Table \ref{tbl:order_within_Dwin_block}. Though all combinations exhibit similar performance patterns, employing depthwise, horizontal, and vertical attentions as the consequtive attentions exhibits more optimal performance. So we fixed this arrangement as the default setting for the NDA layer.  
This is likely due to the fact that at deeper stages of the DwinFormer having a hierarchical structure, it receives input features from the previous stage that are progressively attended in all directions. 
Later, we employ various context aggregators incorporating standard self-attention \cite{dosovitskiy2020image}  as GSA, Swin Transformer \cite{liu2021swin}, CSwin Transformer \cite{cswin}, and the proposed Dwin block at different stages in our network, as shown in Table \ref{ablation}, and observe that employing our hybrid method reflects better dice and HD95 scores. 

\vspace{-0.3cm}
\section{Conclusion}
\vspace{-0.2cm}
We propose DwinFormer to learn the local and global dependencies using  Dwin, and GSA blocks for global feature encoding for better medical image segmentation. The focus of our design is to propose an NDA that progressively increases the receptive field in horizontal, vertical, and depthwise directions as well as a CDA to encode local contextual information for the attention computation.  
Experimental study reveals that our approach provides favorable segmentation results over the multi-organ Synapse dataset and cell HMS dataset.
\\

\noindent\textbf{Compliance with Ethical Standards:} This research study was conducted retrospectively using human subject data made available in open access by the  Multi-Atlas Abdomen Labelling Challenge MICCAI 2015 \cite{landman2015miccai} and zebrafish cell HMS dataset which is an open-source dataset provided by the Department of Systems Biology at Harvard Medical School \cite{cellseg}. For both these datasets, ethical approval was not required as confirmed by the license attached with the open access data. \\

\noindent\textbf{Acknowledgement:}
This work is partially supported by the MBZUAI-WIS Joint Program for AI Research (Project grant number- WIS P008)

\bibliographystyle{IEEEbib}
\bibliography{strings,refs}

\begin{thebibliography}{10}

\bibitem{nnunet}
Fabian Isensee, Jens Petersen, Andre Klein, David Zimmerer, Paul~F. Jaeger, Simon Kohl, Jakob Wasserthal, Gregor Koehler, Tobias Norajitra, Sebastian Wirkert, and Klaus~H. Maier-Hein,
\newblock ``nnu-net: Self-adapting framework for u-net-based medical image segmentation,'' 2018.

\bibitem{a}
Mustansar Fiaz, Moein Heidari, Rao~Muhammad Anwer, and Hisham Cholakkal,
\newblock ``Sa2-net: Scale-aware attention network for microscopic image segmentation,'' 2023.

\bibitem{b}
Daniya~Najiha Abdul~Kareem, Mustansar Fiaz, Noa Novershtern, Jacob Hanna, and Hisham Cholakkal,
\newblock ``Improving 3d medical image segmentation at boundary regions using local self-attention and global volume mixing,''
\newblock {\em IEEE Transactions on Artificial Intelligence}, pp. 1--12, 2023.

\bibitem{ronneberger2015u}
Olaf Ronneberger, Philipp Fischer, and Thomas Brox,
\newblock ``U-net: Convolutional networks for biomedical image segmentation,''
\newblock in {\em International Conference on Medical image computing and computer-assisted intervention}. Springer, 2015, pp. 234--241.

\bibitem{3dunet}
Özgün Çiçek, Ahmed Abdulkadir, Soeren Lienkamp, Thomas Brox, and Olaf Ronneberger,
\newblock ``3d u-net: Learning dense volumetric segmentation from sparse annotation,''
\newblock 06 2016.

\bibitem{zhang2020inter}
Jianpeng Zhang, Yutong Xie, Yan Wang, and Yong Xia,
\newblock ``Inter-slice context residual learning for 3d medical image segmentation,''
\newblock {\em IEEE Transactions on Medical Imaging}, vol. 40, no. 2, pp. 661--672, 2020.

\bibitem{gridach2021pydinet}
Mourad Gridach,
\newblock ``Pydinet: Pyramid dilated network for medical image segmentation,''
\newblock {\em Neural networks}, vol. 140, pp. 274--281, 2021.

\bibitem{wang2019volumetric}
Xudong Wang, Shizhong Han, Yunqiang Chen, Dashan Gao, and Nuno Vasconcelos,
\newblock ``Volumetric attention for 3d medical image segmentation and detection,''
\newblock in {\em Medical Image Computing and Computer Assisted Intervention--MICCAI 2019: 22nd International Conference, Shenzhen, China, October 13--17, 2019, Proceedings, Part VI 22}. Springer, 2019, pp. 175--184.

\bibitem{fang2020spatial}
Wenhao Fang and Xian-hua Han,
\newblock ``Spatial and channel attention modulated network for medical image segmentation,''
\newblock in {\em Proceedings of the Asian Conference on Computer Vision}, 2020.

\bibitem{transunet}
Jieneng Chen, Yongyi Lu, Qihang Yu, Xiangde Luo, Ehsan Adeli, Yan Wang, Le~Lu, Alan~L. Yuille, and Yuyin Zhou,
\newblock ``Transunet: Transformers make strong encoders for medical image segmentation,'' 2021.

\bibitem{nnformer}
Hong-Yu Zhou, Jiansen Guo, Yinghao Zhang, Lequan Yu, Liansheng Wang, and Yizhou Yu,
\newblock ``nnformer: Interleaved transformer for volumetric segmentation,'' 2021.

\bibitem{liu2021swin}
Ze~Liu, Yutong Lin, and et~al.,
\newblock ``Swin transformer: Hierarchical vision transformer using shifted windows,''
\newblock in {\em Proceedings of the IEEE/CVF International Conference on Computer Vision}, 2021, pp. 10012--10022.

\bibitem{dosovitskiy2020image}
Alexey Dosovitskiy, Lucas Beyer, et~al.,
\newblock ``An image is worth 16x16 words: Transformers for image recognition at scale,''
\newblock {\em arXiv preprint arXiv:2010.11929}, 2020.

\bibitem{swin-unet}
Hu~Cao, Yueyue Wang, Joy Chen, Dongsheng Jiang, Xiaopeng Zhang, Qi~Tian, and Manning Wang,
\newblock ``Swin-unet: Unet-like pure transformer for medical image segmentation,'' 2021.

\bibitem{levit}
Guoping Xu, Xingrong Wu, Xuan Zhang, and Xinwei He,
\newblock ``Levit-unet: Make faster encoders with transformer for medical image segmentation,'' 2021.

\bibitem{miss}
Xiaohong Huang, Zhifang Deng, Dandan Li, and Xueguang Yuan,
\newblock ``Missformer: An effective medical image segmentation transformer,'' 2021.

\bibitem{unetr}
Ali Hatamizadeh, Yucheng Tang, Vishwesh Nath, Dong Yang, Andriy Myronenko, Bennett Landman, Holger Roth, and Daguang Xu,
\newblock ``Unetr: Transformers for 3d medical image segmentation,'' 2021.

\bibitem{deeplab}
Huiyu Wang, Yukun Zhu, Bradley Green, Hartwig Adam, Alan~L. Yuille, and Liang{-}Chieh Chen,
\newblock ``Axial-deeplab: Stand-alone axial-attention for panoptic segmentation,''
\newblock {\em CoRR}, vol. abs/2003.07853, 2020.

\bibitem{cellseg}
Andong Wang, Qi~Zhang, Yang Han, Sean Megason, Sahand Hormoz, Kishore~R Mosaliganti, Jacqueline~CK Lam, and Victor~OK Li,
\newblock ``A novel deep learning-based 3d cell segmentation framework for future image-based disease detection,''
\newblock {\em Scientific Reports}, vol. 12, no. 1, pp. 1--15, 2022.

\bibitem{landman2015miccai}
Bennett Landman and et~al.,
\newblock ``Miccai multi-atlas labeling beyond the cranial vault--workshop and challenge,''
\newblock in {\em Proc. MICCAI Multi-Atlas Labeling Beyond Cranial Vault—Workshop Challenge}, 2015, vol.~5, p.~12.

\bibitem{cswin}
Xiaoyi Dong, Jianmin Bao, Dongdong Chen, Weiming Zhang, Nenghai Yu, Lu~Yuan, Dong Chen, and Baining Guo,
\newblock ``Cswin transformer: A general vision transformer backbone with cross-shaped windows,'' 2022.

\end{thebibliography}

\end{document}